\documentstyle[12pt]{article}

\def\be{\begin{equation}}
\def\ee{\end{equation}}
\def\bea{\begin{eqnarray}}
\def\eea{\end{eqnarray}}

\def\dt#1{\on{\hbox{\bf .}}{#1}}                
\def\Dot#1{\dt{#1}}
\def\IR{\relax{\rm I\kern-.18em R}}
\def\binomial#1#2{\left(\,{\buildrel 
{\raise4pt\hbox{$\displaystyle{#1}$}}\over
{\raise-6pt\hbox{$\displaystyle{#2}$}}}\,\right)}

\def\[{\lfloor{\hskip 0.35pt}\!\!\!\lceil}
\def\]{\rfloor{\hskip 0.35pt}\!\!\!\rceil}

\newcommand{\AmS}{{\protect\the\textfont2
  A\kern-.1667em\lower.5ex\hbox{M}\kern-.125emS}}

\catcode`@=11
\def\un#1{\relax\ifmmode\@@underline#1\else
        $\@@underline{\hbox{#1}}$\relax\fi}
\catcode`@=12

\def\fracm#1#2{\hbox{\large{${\frac{{#1}}{{#2}}}$}}}

\def\ad{{\kern0.5pt
                   \alpha \kern-5.05pt
\raise5.8pt\hbox{$\textstyle.$}\kern
0.5pt}}

\def\Dot#1{{\kern0.5pt
     {#1} \kern-5.05pt \raise5.8pt\hbox{$\textstyle.$}\kern
0.5pt}}

\def\a{\alpha}
\def\b{\beta}

\def\d{\delta}
\def\e{\epsilon}

\def\g{\gamma}

\def\k{\kappa}
\def\l{\lambda}
\def\m{\mu}

\def\o{\omega}

\def\q{\theta}

\def\D{\Delta}

\def\O{\Omega}

\def\bo{{\raise.15ex\hbox{\large$\Box$}}}               
\def\pa{\partial}                                       
\def\TH{{\raise.2ex\hbox{$\displaystyle \bigodot$}\mskip-4.7mu \llap H
\;}}
\def\face{{\raise.2ex\hbox{$\displaystyle \bigodot$}\mskip-2.2mu \llap
{$\ddot
        \smile$}}}                                      

\def\Tilde#1{\widetilde{#1}}                    
\def\Hat#1{\widehat{#1}}                        
\def\Bar#1{\overline{#1}}                       
\def\leftrightarrowfill{$\mathsurround=0pt \mathord\leftarrow \mkern-6mu
        \cleaders\hbox{$\mkern-2mu \mathord- \mkern-2mu$}\hfill
        \mkern-6mu \mathord\rightarrow$}
\def\dvec#1{\vbox{\ialign{##\crcr
        \leftrightarrowfill\crcr\noalign{\kern-1pt\nointerlineskip}
        $\hfil\displaystyle{#1}\hfil$\crcr}}}           
\def\dt#1{{\buildrel {\hbox{\LARGE .}} \over {#1}}}     

\def\fracm#1#2{\hbox{\large{${\frac{{#1}}{{#2}}}$}}}
\def\frac#1#2{{\textstyle{#1\over\vphantom2\smash{\raise.20ex
        \hbox{$\scriptstyle{#2}$}}}}}                   
\def\sfrac#1#2{{\vphantom1\smash{\lower.5ex\hbox{\small$#1$}}\over
        \vphantom1\smash{\raise.4ex\hbox{\small$#2$}}}} 
\def\bfrac#1#2{{\vphantom1\smash{\lower.5ex\hbox{$#1$}}\over
        \vphantom1\smash{\raise.3ex\hbox{$#2$}}}}       
\def\afrac#1#2{{\vphantom1\smash{\lower.5ex\hbox{$#1$}}\over#2}}    
\def\on#1#2{\mathop{\null#2}\limits^{#1}}               

\newskip\humongous \humongous=0pt plus 1000pt minus 1000pt
\def\caja{\mathsurround=0pt}
\def\eqalign#1{\,\vcenter{\openup2\jot \caja
        \ialign{\strut \hfil$\displaystyle{##}$&$
        \displaystyle{{}##}$\hfil\crcr#1\crcr}}\,}
\newif\ifdtup

  \def\pp{{\mathchoice
          {
              \kern 1pt%
              \raise 1pt
              \vbox{\hrule width5pt height0.4pt depth0pt
                    \kern -2pt
                    \hbox{\kern 2.3pt
                          \vrule width0.4pt height6pt depth0pt
                          }
                    \kern -2pt
                    \hrule width5pt height0.4pt depth0pt}%
                    \kern 1pt
           }
            {
              \kern 1pt%
              \raise 1pt
              \vbox{\hrule width4.3pt height0.4pt depth0pt
                    \kern -1.8pt
                    \hbox{\kern 1.95pt
                          \vrule width0.4pt height5.4pt depth0pt
                          }
                    \kern -1.8pt
                    \hrule width4.3pt height0.4pt depth0pt}%
                    \kern 1pt
            }
            {
              \kern 0.5pt%
              \raise 1pt
              \vbox{\hrule width4.0pt height0.3pt depth0pt
                    \kern -1.9pt  
                    \hbox{\kern 1.85pt
                          \vrule width0.3pt height5.7pt depth0pt
                          }
                    \kern -1.9pt
                    \hrule width4.0pt height0.3pt depth0pt}%
                    \kern 0.5pt
            }
            {
              \kern 0.5pt%
              \raise 1pt
              \vbox{\hrule width3.6pt height0.3pt depth0pt
                    \kern -1.5pt
                    \hbox{\kern 1.65pt
                          \vrule width0.3pt height4.5pt depth0pt
                          }
                    \kern -1.5pt
                    \hrule width3.6pt height0.3pt depth0pt}%
                    \kern 0.5pt
            }
        }}

  \def\mm{{\mathchoice
                       {
                             \kern 1pt
               \raise 1pt    \vbox{\hrule width5pt height0.4pt depth0pt
                                  \kern 2pt
                                  \hrule width5pt height0.4pt depth0pt}
                             \kern 1pt}
                       {
                            \kern 1pt
               \raise 1pt \vbox{\hrule width4.3pt height0.4pt depth0pt
                                  \kern 1.8pt
                                  \hrule width4.3pt height0.4pt depth0pt}
                             \kern 1pt}
                       {
                            \kern 0.5pt
               \raise 1pt
                            \vbox{\hrule width4.0pt height0.3pt depth0pt
                                  \kern 1.9pt
                                  \hrule width4.0pt height0.3pt depth0pt}
                            \kern 1pt}
                       {
                           \kern 0.5pt
             \raise 1pt  \vbox{\hrule width3.6pt height0.3pt depth0pt
                                  \kern 1.5pt
                                  \hrule width3.6pt height0.3pt depth0pt}
                           \kern 0.5pt}
                       }}

\def\pd{{\kern0.5pt
                   + \kern-5.05pt \raise5.8pt\hbox{$\textstyle.$}\kern
0.5pt}}

\def\pmd{{\kern0.5pt
                  \pm \kern-5.05pt \raise6.3pt\hbox{$\textstyle.$}\kern1.5pt}}

\def\md{{\mathchoice
   {
      {{\kern 1pt - \kern-6.2pt \raise5pt\hbox{$\textstyle.$}\kern 1pt}}}
    {
      {{\kern 1pt - \kern-6.2pt \raise5pt\hbox{$\textstyle.$}\kern 1pt}}}
    {
      {\kern0.5pt - \kern-5.05pt \raise3.4pt\hbox{$\textstyle.$}\kern0.5pt}}
    {
      {\kern0.5pt - \kern-5.05pt \raise3.4pt\hbox{$\textstyle.$}\kern0.5pt}}}}

\def\ad{{\dot{\alpha}}}

\def\pp{{\mathchoice
          {
              \kern 1pt%
              \raise 1pt
              \vbox{\hrule width5pt height0.4pt depth0pt
                    \kern -2pt
                    \hbox{\kern 2.3pt
                          \vrule width0.4pt height6pt depth0pt
                          }
                    \kern -2pt
                    \hrule width5pt height0.4pt depth0pt}%
                    \kern 1pt
           }
            {
              \kern 1pt%
              \raise 1pt
              \vbox{\hrule width4.3pt height0.4pt depth0pt
                    \kern -1.8pt
                    \hbox{\kern 1.95pt
                          \vrule width0.4pt height5.4pt depth0pt
                          }
                    \kern -1.8pt
                    \hrule width4.3pt height0.4pt depth0pt}%
                    \kern 1pt
            }
            {
              \kern 0.5pt%
              \raise 1pt
              \vbox{\hrule width4.0pt height0.3pt depth0pt
                    \kern -1.9pt  
                    \hbox{\kern 1.85pt
                          \vrule width0.3pt height5.7pt depth0pt
                          }
                    \kern -1.9pt
                    \hrule width4.0pt height0.3pt depth0pt}%
                    \kern 0.5pt
            }
            {
              \kern 0.5pt%
              \raise 1pt
              \vbox{\hrule width3.6pt height0.3pt depth0pt
                    \kern -1.5pt
                    \hbox{\kern 1.65pt
                          \vrule width0.3pt height4.5pt depth0pt
                          }
                    \kern -1.5pt
                    \hrule width3.6pt height0.3pt depth0pt}%
                    \kern 0.5pt
            }
        }}

  \def\mm{{\mathchoice
                       {
                             \kern 1pt
               \raise 1pt    \vbox{\hrule width5pt height0.4pt depth0pt
                                  \kern 2pt
                                  \hrule width5pt height0.4pt depth0pt}
                             \kern 1pt}
                       {
                            \kern 1pt
               \raise 1pt \vbox{\hrule width4.3pt height0.4pt depth0pt
                                  \kern 1.8pt
                                  \hrule width4.3pt height0.4pt depth0pt}
                             \kern 1pt}
                       {
                            \kern 0.5pt
               \raise 1pt
                            \vbox{\hrule width4.0pt height0.3pt depth0pt
                                  \kern 1.9pt
                                  \hrule width4.0pt height0.3pt depth0pt}
                            \kern 1pt}
                       {
                           \kern 0.5pt
             \raise 1pt  \vbox{\hrule width3.6pt height0.3pt depth0pt
                                  \kern 1.5pt
                                  \hrule width3.6pt height0.3pt depth0pt}
                           \kern 0.5pt}
                       }}

\def\pd{{\kern0.5pt
                   + \kern-5.05pt \raise5.8pt\hbox{$\textstyle.$}\kern
0.5pt}}

\def\pmd{{\kern0.5pt
                  \pm \kern-5.05pt \raise6.3pt\hbox{$\textstyle.$}\kern1.5pt}}

\def\md{{\mathchoice
   {
      {{\kern 1pt - \kern-6.2pt \raise5pt\hbox{$\textstyle.$}\kern 1pt}}}
    {
      {{\kern 1pt - \kern-6.2pt \raise5pt\hbox{$\textstyle.$}\kern 1pt}}}
    {
      {\kern0.5pt - \kern-5.05pt \raise3.4pt\hbox{$\textstyle.$}\kern0.5pt}}
    {
      {\kern0.5pt - \kern-5.05pt \raise3.4pt\hbox{$\textstyle.$}\kern0.5pt}}}}

\def\dslash{\not{\hbox{\kern-2pt $\partial$}}}
\def\Dslash{\not{\hbox{\kern-4pt $D$}}}
\def\pslash{\not{\hbox{\kern-2.3pt $p$}}}
 \newtoks\slashfraction
 \slashfraction={.13}
 \def\slash#1{\setbox0\hbox{$ #1 $}
 \setbox0\hbox to \the\slashfraction\wd0{\hss \box0}/\box0 }
 
\font\ro=cmsy10                          
\def\kcr{{\hbox{\ro \char'170}}}                
\def\ktl{{\hbox{\ro \char'170}}}        
\def\ktr{{\hbox{\ro \char'170}}}        
\def\kbl{{\hbox{\ro \char'170}}}        
\def\kbr{{\hbox{\ro \char'170}}}        

\def\plpl{\raise-2pt\hbox{$\raise3pt\hbox{$_+$}\hskip-6.67pt\raise0.0pt
\hbox{$^+$}\hskip 0.01pt$}}
\def\mimi{\raise-2pt\hbox{$\raise3pt\hbox{$_-$}\hskip-6.67pt\raise0.0pt
\hbox{$^-$}\hskip 0.01pt$}} 

\def\bo{{\raise.15ex\hbox{\large$\Box$}}}               
\def\pa{\partial}                                       
\def\TH{{\raise.2ex\hbox{$\displaystyle \bigodot$}\mskip-4.7mu \llap H \;}}
\def\face{{\raise.2ex\hbox{$\displaystyle \bigodot$}\mskip-2.2mu \llap {$\ddot
        \smile$}}}                                      

\def\Tilde#1{\widetilde{#1}}                    
\def\Hat#1{\widehat{#1}}                        
\def\Bar#1{\overline{#1}}                       
\def\leftrightarrowfill{$\mathsurround=0pt \mathord\leftarrow \mkern-6mu
        \cleaders\hbox{$\mkern-2mu \mathord- \mkern-2mu$}\hfill
        \mkern-6mu \mathord\rightarrow$}
\def\dvec#1{\vbox{\ialign{##\crcr
        \leftrightarrowfill\crcr\noalign{\kern-1pt\nointerlineskip}
        $\hfil\displaystyle{#1}\hfil$\crcr}}}           
\def\dt#1{{\buildrel {\hbox{\LARGE .}} \over {#1}}}     

\def\fracm#1#2{\hbox{\large{${\frac{{#1}}{{#2}}}$}}}
\def\frac#1#2{{\textstyle{#1\over\vphantom2\smash{\raise.20ex
        \hbox{$\scriptstyle{#2}$}}}}}                   
\def\sfrac#1#2{{\vphantom1\smash{\lower.5ex\hbox{\small$#1$}}\over
        \vphantom1\smash{\raise.4ex\hbox{\small$#2$}}}} 
\def\bfrac#1#2{{\vphantom1\smash{\lower.5ex\hbox{$#1$}}\over
        \vphantom1\smash{\raise.3ex\hbox{$#2$}}}}       
\def\afrac#1#2{{\vphantom1\smash{\lower.5ex\hbox{$#1$}}\over#2}}    
\def\on#1#2{\mathop{\null#2}\limits^{#1}}               

\parskip=\medskipamount                 
\thicklines                         

\def\oldheadpic{                                
        \setlength{\unitlength}{.4mm}
        \thinlines
        \par
        \begin{picture}(349,16)
        \put(325,16){\line(1,0){4}}
        \put(330,16){\line(1,0){4}}
        \put(340,16){\line(1,0){4}}
        \put(335,0){\line(1,0){4}}
        \put(340,0){\line(1,0){4}}
        \put(345,0){\line(1,0){4}}
        \put(329,0){\line(0,1){16}}
        \put(330,0){\line(0,1){16}}
        \put(339,0){\line(0,1){16}}
        \put(340,0){\line(0,1){16}}
        \put(344,0){\line(0,1){16}}
        \put(345,0){\line(0,1){16}}
        \put(329,16){\oval(8,32)[bl]}
        \put(330,16){\oval(8,32)[br]}
        \put(339,0){\oval(8,32)[tl]}
        \put(345,0){\oval(8,32)[tr]}
        \end{picture}
        \par
        \thicklines
        \vskip.2in}
\def\oldtitle#1#2#3#4{\oldheadpic\begin{center}\vglue.5in{\large\bf #1}\\[.6in]
        {#2}\\[.1in] {\it Department of Physics and Astronomy}\\
        {\it University of Maryland, College Park, MD 20742}\\[.6in]
        Physics Publication \#{#3}\\ {#4}\\[1.5in] {\bf ABSTRACT}\\[.1in]
        \end{center} \begin{quotation}}                 
\def\oldTitle#1#2#3#4#5#6#7{\oldheadpic\begin{center} \vglue .4in
        {\large\bf #1}\\[.4in]
        {#2}\\[.1in] {\it Department of Physics and Astronomy}\\
        {\it University of Maryland, College Park, MD 20742}\\[.1in]
        {#3}\\[.1in] {\it {#4}}\\ {\it {#5}}\\[.4in]
        Physics Publication \#{#6}\\ {#7}\\[.5in] {\bf ABSTRACT}\\[.1in]
        \end{center} \begin{quotation}}                 
\def\border{                                            
        \setlength{\unitlength}{1mm}
        \newcount\xco
        \newcount\yco
        \xco=-21
        \yco=12
        \begin{picture}(140,0)
        \put(\xco,\yco){$\ktl$}
        \advance\yco by-1
        {\loop
        \put(\xco,\yco){$\kcr$}
        \advance\yco by-2
        \ifnum\yco>-240
        \repeat
        \put(\xco,\yco){$\kbl$}}
        \xco=158
        \yco=12
        \put(\xco,\yco){$\ktr$}
        \advance\yco by-1
        {\loop
        \put(\xco,\yco){$\kcr$}
        \advance\yco by-2
        \ifnum\yco>-240
        \repeat
        \put(\xco,\yco){$\kbr$}}
        \put(-20,13){\tiny University of Maryland Elementary Particle
Physics University of Maryland Elementary Particle Physics University of
Maryland Elementary Particle Physics}
        \put(-20,-241.5){\tiny University of Maryland Elementary
Particle Physics University of Maryland Elementary Particle Physics
University of Maryland Elementary Particle Physics}
        \end{picture}
        \par\vskip-8mm}
\def\bordero{                                           
        \setlength{\unitlength}{1mm}
        \newcount\xco
        \newcount\yco
        \xco=-31
        \yco=12
        \begin{picture}(140,0)
        \put(\xco,\yco){$\ktl$}
        \advance\yco by-1
        {\loop
        \put(\xco,\yco){$\kclr}
        \advance\yco by-2
        \ifnum\yco>-240
        \repeat
        \put(\xco,\yco){$\kbl$}}
        \xco=151
        \yco=12
        \put(\xco,\yco){$\ktr$}
        \advance\yco by-1
        {\loop
        \put(\xco,\yco){$\kcr$}
        \advance\yco by-2
        \ifnum\yco>-240
        \repeat
        \put(\xco,\yco){$\kbr$}}
        \put(-20,12){\ooo bacdefghidfghghdhededbihdgdfdfhhdheidhdhebaaahjhhdahba

hgdedge
   hgfdiehhgdigicba}
        \put(-20,-241.5){\ooo ababaighefdbfghgeahgdfgafagihdidihiidhiagfedhadbfd

ecdcdfa
   gdcbhaddhbgfchbgfdacfediacbabab}
        \end{picture}
        \par\vskip-8mm}
\def\headpic{                                           
        \indent
        \setlength{\unitlength}{.4mm}
        \thinlines
        \par
        \begin{picture}(29,16)
        \put(165,16){\line(1,0){4}}
        \put(170,16){\line(1,0){4}}
        \put(180,16){\line(1,0){4}}
        \put(175,0){\line(1,0){4}}
        \put(180,0){\line(1,0){4}}
        \put(185,0){\line(1,0){4}}
        \put(169,0){\line(0,1){16}}
        \put(170,0){\line(0,1){16}}
        \put(179,0){\line(0,1){16}}
        \put(180,0){\line(0,1){16}}
        \put(184,0){\line(0,1){16}}
        \put(185,0){\line(0,1){16}}
        \put(169,16){\oval(8,32)[bl]}
        \put(170,16){\oval(8,32)[br]}
        \put(179,0){\oval(8,32)[tl]}
        \put(185,0){\oval(8,32)[tr]}
        \end{picture}
        \par\vskip-6.5mm
        \thicklines}
\def\title#1#2#3#4{\border\headpic {\hbox to\hsize{#4 \hfill UMDEPP #3}}\par
        \begin{center} \vglue .5in {\large\bf #1}\\[.6in]
        {#2}\\[.1in] {\it Department of Physics and Astronomy}\\
        {\it University of Maryland, College Park, MD 20742}\\[1.5in]
        {\bf ABSTRACT}\\[.1in] \end{center} \begin{quotation}}  
\def\Title#1#2#3#4#5#6#7{\border\headpic
        {\hbox to\hsize{#7 \hfill UMDEPP #6}}\par
        \begin{center} \vglue .4in {\large\bf #1}\\[.4in]
        {#2}\\[.1in] {\it Department of Physics and Astronomy}\\
        {\it University of Maryland, College Park, MD 20742}\\[.1in]
        {#3}\\[.1in] {\it {#4}}\\ {\it {#5}}\\[.5in] {\bf ABSTRACT}\\[.1in]
        \end{center} \begin{quotation}}                 
\def\endtitle{\end{quotation}\newpage}                  

\def\qd{{\kern0.5pt
                   q \kern-5.05pt \raise5.8pt\hbox{$\textstyle.$}\kern
0.5pt}}

\begin{document}

\def\dt#1{\on{\hbox{\bf .}}{#1}}                
\def\Dot#1{\dt{#1}}

\def\gfrac#1#2{\frac {\scriptstyle{#1}}
        {\mbox{\raisebox{-.6ex}{$\scriptstyle{#2}$}}}}
\def\gg{{\hbox{\sc g}}}
\border\headpic {\hbox to\hsize{January 2001 \hfill
{UMDEPP 00-032}}}
\vskip-0.12in
\par \hfill {hep-th/0101037\,}
\vskip-0.12in
\par \hfill {(Revised Version)}
\par
\setlength{\oddsidemargin}{0.3in}
\setlength{\evensidemargin}{-0.3in}
\begin{center}
\vglue .1in
{\Large\bf Deliberations on 11D \\
Superspace for the \\ \vglue .15in
M-Theory Effective Action\footnote{
Supported in part by National Science Foundation Grant 
PHY-98-02551.}  }
\\[.25in]
S.\ James Gates, Jr.\footnote{gatess@wam.umd.edu}
and Hitoshi Nishino\footnote{nishino@katherine.physics.umd.edu}
\\[0.06in]
{\it Department of Physics\\ University of Maryland\\ 
College Park, MD 20742-4111 USA}
\\[0.4in]

{\bf ABSTRACT}\\[.01in]
\end{center}
\begin{quotation}
{In relation to the superspace modifications of 11D supergeometry
required to describe the M-theory low-energy effective action, we 
present an analysis of infinitesimal supergravity fluctuations about 
the flat superspace limit.   Our investigation confirms Howe's 
interpretation of our previous Bianchi identity analysis.  However, 
the analysis also shows that should 11D supergravity obey the rules 
of other off-shell supergravity theories, the complete M-theory 
corrections will necessarily excite our previously anticipated 
spin-1/2 engineering dimension-1/2 spinor auxiliary multiplet 
superfield. The analysis of fluctuations yields more evidence 
that Howe's 1997 theorem is specious when applied to Poincar\' e 
supergravity or 11D supergravity/M-theory.  We end by commenting 
upon recent advances in this area.}  

\bigskip\bigskip

${~~~}$ \newline
\leftline{PACS: 03.70.+k, 11.30.Rd, 04.65.+e} 
\vskip-0.30in    
${~~~}$ \newline
\leftline{Keywords: Gauge theories, Supersymmetry,
Supergravity.}

\endtitle

\baselineskip 13.9pt 

\section {Introduction}  

~~~~During the middle eighties, some of our research \cite{G1} {\it 
{initiated}} a new direction of study for superspace geometry, {\it
i.e.,} the  problem of how to describe higher derivative supergravity
theories {\it via} the use of on-shell superspace techniques.  Prior 
to our work \cite{G1,G2},  this area had received {\it {no}} attention 
in the physics literature. We proposed that since low-energy superstring
effective actions naturally  contain a dimensionful parameter $\a'$ (in
our discussions we used a symbol $\g'$ which is proportional to $\a'$) it
should be expected  that supergeometries, associated with superstring and
heterotic string effective actions, should themselves become expansions
in this dimensionful parameter.  Thus was introduced into the literature
the notion of the ``slope parameter expansion'' of supergeometries, with
the work of \cite{G1} providing a proof-in-principle that this technique 
could be viable.  This proof-in-principle, however, was restricted to 
the {\it {dual}} formulation of 10D, $N$ = 1 supergravity.   Our early 
investigations of the dual theory also marked the {\it {initial}} 
discovery of type I closed string/type I fivebrane duality by showing 
that under a duality transformation, the stringy higher mode 
corrections and the stringy quantum loop corrections are exchanged.  

It was not until later work \cite{G3,G4}, wherein the concept of the 
``Lorentz gaugino'' was introduced, before we were able to extend our 
proof-in-principle to the actual Chern-Simons related corrections of 
the low-energy heterotic string effective action. This was possible 
because the ``Lorentz gaugino'' maps problems involving 10D supergravity 
into analogous and often simpler problems involving 10D super Yang-Mills
theory. The second of these works also revealed that the ``Lorentz
gaugino'' takes its simplest form ({\it i.e}., only proportional to the
gravitino curl)  when the superspace constraints utilized correspond to
what is presently  called the ``string frame.''  The discovery of these
special superspace constraints \cite{GNZ} was made well before this
terminology existed, so we simply referred to these as ``improved 
supergeometries'' and showed that they also existed to type-II theories 
\cite{G5}.  The concept of the Lorentz gaugino has also played a key 
role in component discussions of higher derivative supergravity.  A 
perusal of the initial work by Bergshoeff and de Roo \cite{BdR}, shows 
that after the discovery of this concept in superspace, these researchers 
utilized it to elevate the level of understanding of higher curvature 
supergravity at the component level.

Finally in the early nineties \cite{G6}, we returned to this class of 
problems by showing, in some detail, how to modify 11D supergravity 
based on a 1980 conjecture \cite{G80}.   We proposed \cite{G6} that 
this suggestion would be critical for discussions of the 11D
supergeometry appropriate to encode information about the low-energy
effective action of M-theory.   In particular, our assertion implied
the presence of a dimension one-half and spin-1/2 multiplet of
currents.   In fact, we suggested that 4D, $N$ = 2 supergravity and/or 
4D, $N$ = 1 nonminimal supergravity was likely to provide the best 
paradigms to follow in  building the 11D superspace theory.   We shall 
argue in the following that this is precisely what has happened in the 
period since our work.

Shortly after the introduction of our modified 11D supergeometry, there 
appeared a paper by Howe \cite{HOW}.   A ``theorem'' was proposed that 
implied the equations of motion of 11D supergravity/M-theory follow 
{\it {solely}} from constraints on (engineering) dimension zero
superspace torsions.   A misunderstanding of our proposal began his 
criticism of our 1996 work by stating, ``We conclude with some brief 
comments on  a recent paper [9] by Nishino and Gates in which it
is {\it {claimed that  an off-shell extension of eleven-dimensional 
supergravity can be constructed in superspace involving dimension
one-half superfields}}.''  The penultimate paragraph in our work,
however, had stated our actual position, ``For although we believe 
our observation is important, we know of at least {\it {two}}  
arguments that suggest that there must exist {\it {at least one other 
tensor superfield that will be required to have a completely off-shell
formalism}}\footnote{We have italicized parts in this paragraph to
emphasize what was {\it {actually}} written.}.''   

Therefore we never proposed that ``...an  off-shell extension of
eleven-dimensional supergravity can be constructed in superspace
involving a dimension   one-half superfield.'' We clearly stated that
our ${\cal J}_{\a}$ tensor  was to be regarded as {\it {only}} part of
an off-shell theory.  We referred  to an 11D analog of the non-minimal
4D, $N$ = 1 theory \cite{GSW} and containing an  ``eleven dimensional
analog of the $G_{\un a}$'' as being the most likely candidate for 11D
supergravity/M-theory modified supergeometry.  

In the following, we present an analysis based on the study of
infinitesimal 11D supergravity fluctuations about a flat background
that allows us to  clarify our position.  This analysis will show that;
(a.) our 1996 embedding  of ${\cal J}_{\a}$ in the torsion supertensors
was {\it {incorrect}}, (b.)  permits us to find the correct embedding of
${\cal J}_{\a}$ into the d = 1/2  torsion tensors, (c.) provides us with
yet another basis for claiming that the best model for off-shell 11D
supergravity is the non-minimal 4D, $N$ = 1 theory and (d.) provides 
a proof invalidating Howe's 1997 theorem.

Our result will also allow us to suggest modifications to the
``$X$-tensor'' approach given recently \cite{CGNN}.  Parts of these
field strength superfields are shown to be purely ``conventional
constraints'' \cite{GSW,GSWc} whose vanishing simply implies the
algebraic elimination of parts of the linearized 11D vielbein. We also
will give the {\it {first}} explicit supergravity/M-theory currents
in both the ${\cal J}$-tensor and $X$-tensor approaches.

\section {The 11D Vielbein: 43 $\cdot 2^{37}$ Degrees of Freedom}

~~~~In order to better understand 11D supergravity, we believe that it 
is useful to look ``behind the geometrical curtain'' that has been used 
in all previous discussions.  A step toward this is introducing the
following  spinorial 11D vielbein parametrization\footnote{Although we
believe this is  natural parametrization that can be applied to
supergravity \newline ${~~~~~}$ in other dimensions, this has not 
appeared to our knowledge previously as applied to 10D \newline $
{~~~~~}$ and 11D supergravity theories.} 
\be \eqalign{ 
{\rm E}{}_\a & =~ \Psi^{1/2} \, \Big\{ \, \exp \big( \fracm 12 \, \D
    \big) ~ \Big\}_\a{}^\b ~ \Big( \, D_\b ~+~ {\rm H}{}_\b{}^m 
     \, \pa_m \, \Big) ~\equiv~ \Psi^{1/2} \, {\cal N}{}_\a 
    {}^\b ~ \Big( \, 
     D_\b ~+~ {\rm H}{}_\b{}^m \, \pa_m \, \Big) ~~, 
\cr  
\Delta_\a{}^\b & \equiv ~ \Big[ \, i \Psi^a 
    \g_a \, + \, \fracm 1 2 \Psi^{a \, b} \g_{a \, b} 
     \,+\, i \fracm 1 6 
    \Psi^{ \[ 3 \] }  \g_{\[ 3 \]} {~~} 
     +\,\fracm 1{24} \Psi^{\[ 4 \]} \g_{\[ 4 \]} \,+\, i \fracm 1{120}
   \Psi^{\[ 5 \]} \g_{\[ 5 \]} \, \Big]_\a{}^\b   ~. 
\label{eq:01} }   \ee 
The factor $ {\cal N}{}_{\a} {}^{\b}$ above is an element of the
SL(32,\IR) group since det$\,({\cal N}{}_{\a} {}^{\b}) = 1$ and $({\cal 
N}{}_{\a}{}^{\b})^* = {\cal N}{}_{\a} {}^{\b}$.  The superfields
$\Psi$, $\Psi^{a}$ through $\Psi^{\[5\]}$ are forms with respect to 
the 11D tangent Lorentz space which allows us to easily count the number  
of their independent components.   
\vspace{0.1cm}
\begin{center}
\footnotesize
\begin{tabular}{|c|c|c|}\hline
$p$ & $p -{\rm {form}}$ & ${\rm {Form}}~{\rm {Dimension}}:~
D_p = \binomial{11} p $ 
\\ \hline
$ ~~0 ~~$ & $ ~~ \Psi~~$ & $ ~D_0 \,= \, 1 \,~~\,~$ \\ \hline
$ ~~1 ~~$ & $ ~~ \Psi^a~~$ & $ ~ D_1 \,= \, 11 ~\,~$ \\ \hline
$ ~~2 ~~$ & $ ~~ \Psi^{\[2\]}~~$ & $ ~ D_2 \,= \, 55 ~\,~$ \\ \hline
$ ~~3 ~~$ & $ ~~ \Psi^{\[3\]}~~$ & $ ~ D_3 \,= \, 165 ~$ \\ \hline
$ ~~4 ~~$ & $ ~~ \Psi^{\[4\]}~~$ & $ ~ D_4 \,= \, 330 ~$ \\ \hline
$ ~~5 ~~$ & $ ~~ \Psi^{\[5\]}~~$ & $ ~ D_5 \,= \, 462 ~$ \\ \hline
\end{tabular}
\end{center}
\begin{center}
{Table 1: Holonomy p-forms of the 11D Vielbein}
\end{center}
The form (but {\it {not}} super-form) superfields can be called holonomy
$p$-forms because they define the most general  holonomy group acting {\it
{solely}} on the spinorial frame.  The remaining  superfield entering the
vielbein is ${\rm H}{}_{\a}{}^m$ which plays  the analogous role to conformal
prepotentials in lower dimension superspace  supergravity theories.  

While the component  11D vielbein ${\rm e}_a {}^m$ contains only 121
bosonic  d.f.\, the component 3-form contains only 165 bosonic d.f.\
\footnote{ d.f.\ 
$\equiv$ degrees of freedom} and the component gravitino $\psi_a
{}^{\a}$  contains 356 fermionic d.f.\ , the total number of component
d.f.\ in ${\rm E}{}_{\a}{}^M$ is astronomical with  $ 43 \cdot \, 
2^{36}$ bosonic d.f.\ and the same number of fermionic d.f.!  These are
distributed according to the following table,
\vspace{0.1cm}
\begin{center}
\footnotesize
\begin{tabular}{|c|c|c|}\hline
${\rm {Superfield}}$ & ${\rm {Bosonic}}-{\rm {d.\ f.\ }}$ & 
${\rm {Fermionic}}-{\rm {d.\ f.\ }}$ 
\\ \hline
$ ~~ \Psi~~$ & $ ~{\rm d}_S  \,~~\,~$ & $ ~  {\rm d}_S ~$ \\ \hline
$ ~~ \Psi^a~~$ & $ ~11 \cdot   {\rm d}_S ~\,~$ & $ ~11 \cdot   {\rm d}_S~$ \\
\hline
$ ~~ \Psi^{\[2\]}~~$ & $ ~ 55  \cdot   {\rm d}_S~\,~$ & 
     $ ~ 55 \cdot {\rm d}_S~$ \\
\hline
$ ~~ \Psi^{\[3\]}~~$ & $ ~165  \cdot   {\rm d}_S~$ & 
     $ ~165  \cdot{\rm d}_S~$ \\ \hline
$ ~~ \Psi^{\[4\]}~~$ & $ ~330  \cdot   {\rm d}_S~$ & 
      $ ~330 \cdot   {\rm d}_S~$ \\ \hline
$ ~~ \Psi^{\[5\]}~~$ & $ ~462  \cdot   {\rm d}_S~$ & 
     $ ~462  \cdot   {\rm d}_S~$ \\ \hline
$ ~~ {\rm H}{}_{\b}{}^m 
~~$ & $ ~ 352 \cdot  {\rm d}_S ~$ & 
     $ ~352 \cdot  {\rm d}_S ~$ \\ \hline
     $ ~~ {\rm E}{}_{\a} 
~~$ & $ ~ 1,376 \cdot   {\rm d}_S~$ & 
      $ ~ 1,376 \cdot   {\rm d}_S ~$ \\ \hline
\end{tabular}
\end{center}
\begin{center}
{Table 2: Degrees of Freedom in Semi-prepotentials}
\end{center}
where ${\rm d}_S$ = $2^{31}$  = 2,147,483,648.

Whatever choice of parametrization of the spinorial vielbein is made,
following  the experience of all previous constructed solutions to
superspace supergravity  constraints, we may define the vectorial frame
through the equation
\be \eqalign{
{\rm E}{}_{a} \, &\equiv\,  i \, \fracm 1{\,32\,} \, (\g_a)^{\a \, 
\b} \, \[\, {\rm E}{}_{\a} ~,~ {\rm E}{}_{\b} \, \} 
~~~. \cr
} \label{eq:02} \ee
This is not just a convenience. Without enforcing such a constraint, the 
superspace formulation does not contain a unique graviton.  Following
the  exact same arguments that were first given many, many years ago
\cite{ConSTRT},  if there are absolutely {\it {no}} constraints imposed
on the superspace frame fields then there exist a Wess-Zumino gauge in
which 
\be  \eqalign{ {~~~~}
{\rm E}{}_{\a}{}^m (\q, \,x) &=~  \fracm 12 \q^{\b} \Big[ \, i(\g^k)_{
\a \, \b}\, {{\rm e}}{}_{k}{}^m (x)  ~+~ \,(\g^{k \, l})_{\a \, \b} \,
{\rm f}_{k\, l} {}^{m}(x)  ~+~i\,(\g^{\[5\]})_{\a \, \b}\, 
{\rm f}_{\[5\]}{}^m (x) \, \Big] \cr 
&~~~~+~ ({\rm {higher}} ~{\rm {order}}~ \q\,\hbox{-terms}) ~~~,\cr 
{\rm E}{}_{a}{}^m (\q, \,x) &=~ {\Hat {\rm e}}{}_{a}{}^m (x) 
    ~+~ ({\rm {higher}} ~{\rm {order}}~\,
     \theta\,\hbox{-terms}) ~~~, }   
\label{eq:03} \ee
and there are seen to be {\it {two}} a priori independent fields, ${{\rm
e}}{}_{a} {}^m (x)$ and ${\Hat {\rm e}}{}_{a}{}^m (x)$, whose
transformation laws make both candidates to be identified with the 
usual component graviton.  It is precisely the role of the condition in
(\ref{eq:02}) to insure that these are, in fact, the same field.  This
is clearly an important physical requirement since all conventional
component formulations possess a unique graviton.

It has been known since the early eighties how to write the solution
to conditions such as that appearing in (\ref{eq:02}).  Firstly, one notices
that the quantities defined by
\be \eqalign{ 
{\Hat {\rm E}}{}_{\a} &\equiv~ D_{\a} ~+~  {\rm H}{}_{\a}{}^d
\pa_d ~ 
~~~,~~~ \cr
{\Hat {\rm E}}{}_{a} &\equiv~ \pa_a ~+~ i \, \fracm 1{\,16\,} 
\, (\g_a)^{\a \, \b} \, [\, D_{\a} {\rm H}{}_{\b}{}^d  ~+~ {\rm H}{}_{
\a}{}^c \, \pa_c  {\rm H}{}_{\b}{}^d \,] \, \pa_d ~ \cr
&\equiv~ \pa_a ~+~ {\rm H}{}_{a}{}^d \, \pa_d ~\equiv~ {\Hat 
{\rm E}}{}_{a}{}^m \pa_m ~~~,
}\label{eq:03A} \ee 
imply the equations,
\be \eqalign{
\[ \, {\Hat {\rm E}}{}_{\a} ~,~ {\Hat {\rm E}}{}_{\b} \, \}  ~=~ [~
i \, (\g^c)_{\a \, \b} ~+~  {\Hat {\rm C}}{}_{\a \b}{}^c ~ ] \, {\Hat 
{\rm E}}{}_c  ~~~,~~~ (\g_a)^{\a \, \b} \, {\Hat {\rm C}}{}_{\a \b}{}^c 
~=~ 0 ~~~,
}\label{eq:03B} \ee 
where an explicit expression for the purely imaginary quantity\footnote{
We use ``superspace conjugation'' in order to define reality properties.
This permits factors of \newline ${~~~~\,}$ $i$ to appear consistently
even though our supercoordinates $(\q^{\a},\, x^a)$ for 11D superspace 
are real. \newline ${~~~~\,}$ A recent pedagogical discussion of this
operation can be found in \cite{TASI}.} ${\Hat {\rm C}}{}_{\a  \b}{}^c$ is
given by
\be 
{\Hat {\rm C}}{}_{\a \b}{}^c ~=~ [\, \d_{( \a}{}^{\g} \d_{\b)}{}^{\d}
~+~ \fracm 1{16} \, (\g^a)_{\a \, \b} \, (\g_a)^{\g \, \d} \, ] 
    \, [\, D_{\g} 
     {\rm H}{}_{\d}{}^m  ~+~ {\rm H}{}_{\g}{}^b \, \pa_b  
     {\rm H}{}_{\d}{}^m \,]  \, {\Hat {\rm E}}{}_m {}^c ~~~.
\label{eq:03C} \ee
The quantity $ {\Hat {\rm E}}{}_m {}^c$ is the inverse to the superfield
$ {\Hat {\rm E}}{}_a {}^m$ that appears in (\ref{eq:03A}).  The general 
calculation of the commutator algebra of ${\Hat {\rm E}}{}_A$ leads to
the remaining relations
\be \eqalign{
\[ \, {\Hat {\rm E}}{}_{\a}  ~,~ {\Hat {\rm E}}{}_b \, \} &=~ 
{\Hat C}{}_{\a b}{}^c \, {\Hat {\rm E}}{}_c  ~~~, ~~~
\[ \, {\Hat {\rm E}}{}_a  ~,~ {\Hat {\rm E}}{}_b \, \} ~=~ 
{\Hat C}{}_{a b}{}^c \, {\Hat {\rm E}}{}_c  ~~~, \cr
} \label{eq:03CC} \ee
and it is seen that {\it {only}} the vectorial frame appears on the
right hand sides of these equations as also is the case in 
(\ref{eq:03B}).  Explicitly we find
\be \eqalign{
{\Hat C}{}_{\a b}{}^c &=~ \big(\, D_{\a}  {\rm H}{}_{b}{}^m 
     ~+~ {\rm H}{}_{\a} 
{}^e \pa_e  {\rm H}{}_{b}{}^m  ~-~ \pa_b  {\rm H}{}_{\a}{}^m 
     ~-~ {\rm H}
{}_{b} {}^e \pa_e  {\rm H}{}_{\a}{}^m \, \big) 
   \, {\Hat {\rm E}}{}_m {}^c  ~~~,  \cr 
{\Hat C}{}_{a b}{}^c &=~ \big(\, \pa_a  {\rm H}{}_{b}{}^m 
    ~+~ {\rm H}{}_{a} 
{}^e \pa_e  {\rm H}{}_{b}{}^m   ~-~ \pa_b  {\rm H}{}_{a}{}^m 
    ~-~ {\rm H}
{}_{b} {}^e \pa_e  {\rm H}{}_{a}{}^m\, \big) \, {\Hat {\rm E}}{}_m {}^c 
    ~~~. \cr
}\label{eq:03CCC} \ee
It is also an interesting fact that 
\be \eqalign{
{\rm sdet}\,({\Hat {\rm E}}{}_A {}^M ) &=~ 
\det\,({\Hat {\rm E}}{}_a {}^m ) ~=~ \det\big(\, \d_a {}^m 
     ~+~  {\rm H}{}_{a}{}^m \,\big) ~~~.
}\label{eq:03D} \ee

The set of superframes ${\Hat {\rm E}}{}_A {}^M $ are not the ones we
require, however.  Comparing (\ref{eq:01}) with the first
line in (\ref{eq:03A}) we see
\be 
{\rm E}{}_{\a} ~=~ \Psi^{1/2} \, {\cal N}_{\a}{}^{\g} \,
{\Hat {\rm E}}{}_{\g}  ~~~, 
\label{eq:03E} \ee
and this can be substituted into (\ref{eq:02}) to derive
\be  \eqalign{
{\rm E}{}_{a} ~=~ &\Psi \, {\cal N}_a {}^b \, {\Hat {\rm E}}{}_b 
     ~+~ i
\, \fracm 1{\,32\,} \, (\g_a)^{\a \, \b} \, (\, {\rm E}{}_{\a} 
    \, \ln \Psi  \,)  \,{\rm E}{}_{\b}  \cr 
&+~ i \, \fracm 1{\,16\,} \, (\g_a)^{\a \, \b} \, [\, {\rm E}{
}_{\a} {\cal N}{}_{\b}{}^{\d} \, ]\,({\cal N}^{-1})_{\d}{}^{\g}
\, {\rm E}{}_{\g}   ~~~, 
}\label{eq:03F} \ee
where the real quantity $ ({\cal N}^{-1})_a {}^b$ is the matrix inverse
to ${\cal N}_a {}^b$ defined by
\be
{\cal N}_a {}^b ~=~ - \fracm 1{32} \, (\g_a)^{\a \, \b}
      \, {\cal N}{
}_{\a}{}^{\g} \, {\cal N}{}_{\b}{}^{\d} \, [~ (\g^b)_{\g \, \d} 
     ~-~  i\,  {\Hat C}{}_{\g \, \d}{}^b ~ ] ~~~.
\label{eq:03G} \ee
The last factor in (\ref{eq:03F}) has an interesting group theoretical
interpretation.  If we denote the exterior differential by $d$, then the 
quantity $\O_{\b}{}^{\g}$ defined by
\be
\O_{\a}{}^{\b} ~\equiv~ [\, d {\cal N}{}_{\a}{}^{\g} \, ] 
     \, ({\cal N}^{-1})_{\g}{}^{\b} ~~~,
\label{eq:03H} \ee
with $\O_{\a}{}^{\a} = 0$ is the right-invariant Maurer-Cartan form of
SL(32,\IR) satisfying the condition $d \O =  \O \wedge \O $.  

The commutator algebra of the set of superframes ${{\rm E}}{}_A {}^M $
is considerably more complicated than that of the set of superframes 
${\Hat {\rm E}}{}_A {}^M $.  Direct calculation reveals
\be \eqalign{
\[\, {\rm E}{}_{\a} ~,~ {\rm E}{}_{\b} \, \} 
     ~=~ {\rm C}{}_{\a \b}{}^{\g}  
\, {\rm E}{}_{\g} ~+~ i \, [~  (\g^c)_{\a \, \b} 
     ~-~ i \,  {\rm C}{}_{\a \b} {}^c  ~ ] \, {\rm E}{}_c   ~~~,
}\label{eq:03B2} \ee 
where the full spinor-spinor anholonomy coefficients take the forms
\be  \eqalign{ {~~~~~~~}
C{}_{\a \, \b}{}^c &=~ i \,\fracm 1{64} \, [\,  (\g^{\[2\]})_{\a \b} 
\, (\g_{\[2\]})^{\k \,\l} ~-~  \fracm 1{60} \,  (\g^{\[5\]})_{\a \b} 
\, (\g_{\[5\]})^{\k \,\l} \,] \, {\cal F}_{\k \l}{}^c   ~~~, 
     {~~~~~~~~~}
}\label{eq:03HA} \ee
\be  \eqalign{ {~~~}
C{}_{\a \, \b}{}^{\g} &=~ \fracm 1{64} \, [\, (\g^{\[2\]})_{\a \b} 
      \, (\g_{\[2\] 
})^{\k \,\g} ~-~ \fracm 1{60} \,  (\g^{\[5\]})_{\a \b} 
     \, (\g_{\[5\]})^{\k \,\g} 
\, ]\,  (\, {\rm  E}{}_{\k} \, \ln \Psi \,)    \cr 
&{~~~~}+~ \fracm 1{\,2,048} \, [\,  (\g^{\[2\]})_{\a \b} 
    \, (\g_{\[2\]})^{\k \,\l}  
~-~ \fracm 1{60} (\g^{\[5\]})_{\a \b} \, (\g_{\[5\]})^{\k \,\l} 
     \, ] \,
{\cal  F}_{\k \l}{}^c \, (\g_c)^{\k \,\g} \, (\, {\rm E}{}_{\k} 
     \, \ln \Psi \,)  \cr 
&{~~~~}+~\fracm 1{32} \, [\, (\g^{\[2\]})_{\a \b} \,
(\g_{\[2\]})^{\k \,\l} ~-~ 
\fracm 1{60} \, (\g^{\[5\]})_{\a \b} \, (\g_{\[5\]})^{\k \,\l} \,] 
    \, (\, {\rm
    E}{}_{\k} \, {\cal N}_{\l} {}^{\d}\,) \, ({\cal N}^{-1})_{\d}{}^{\g}
    \cr  
&{~~~~}+~ \fracm 1{\,1,024} \, (\g^{\[2\]})_{\a \b} \,
    (\g_{\[2\]})^{\k \,\l}  
   \, {\cal F}_{\k \l}{}^c \, (\g_c)^{\k \,\d} 
    \, (\, {\rm E}{}_{\k} \,
   {\cal N}_{\d}{}^{\eta} \,)  ({\cal N}^{-1})_{\eta}{}^{\g} \cr
&{~~~~}-~ \fracm 1{\,61,440} \, (\g^{\[5\]})_{\a \b} 
     \, (\g_{\[5\]})^{\k \,\l}  
    \, {\cal F}_{\k \l}{}^c \, (\g_c)^{\k \,\d} 
   \, (\, {\rm E}{}_{\k} \,
{\cal N}_{\d}{}^{\eta} \,)  ({\cal N}^{-1})_{\eta}{}^{\g}   ~~~,
}\label{eq:03HB} \ee
and the real quantity $ {\cal F}_{\k \l}{}^c $ is defined according to
\be
{\cal F}_{\k \l}{}^c ~\equiv~ {\cal N}{}_{\k}{}^{\g} \, {\cal N}{}_{\l}
{}^{\d} \, [~ (\g^d)_{\g \, \d} \,-\, i \,{\Hat {\rm C}}{}_{\g \d}{}^d
~ ] \,  ({\cal N}{}^{-1})_d {}^c ~~~.
\label{eq:03HC} \ee
Although the complete analogs of all the results in (\ref{eq:03CC}) can
be explicitly calculated, we will defer to some future date such a
presentation. The most important point of the non-linear analysis
completed above is that it proves  that the condition in (\ref{eq:02})
has a solution in terms of
$\Psi_{\[p\]}$  and ${\rm H}{}_{\a}{}^m$ such that {\it {no}}
restrictions are placed on these superfields in spite of the fact that
(\ref{eq:02})  implies
\be 
(\g_b)^{\a \, \b} \, {\rm C}{}_{\a \b}{}^c ~=~ 
(\g_b)^{\a \, \b} \, {\rm C}{}_{\a \b}{}^{\g} ~=~ 0 ~~~.
\label{eq:03J} \ee
Finally, due to the definition (\ref{eq:02}) and its solution discussed
above  we see
\be
{\rm sdet}\,({\rm E}{}_A {}^M ) 
    ~=~ \Psi^{- 5} ~ \det\, ({\cal N}_a {}^b)
     ~ \det\,({\Hat {\rm E}}{}_a {}^m ) ~~~.
\label{eq:03I} \ee

A special gauge choice of the expressions in (\ref{eq:01}) and
(\ref{eq:02})  corresponds to 11D supergeometries which are conformally
related to flat 11D  geometry superspace
\be \eqalign{ {~~~~~}
{\rm E}{}_{\a} ~=~ \Psi^{1/2} \, D_{\a} ~~~,~~~ {\rm E}{}_{a} ~=~  
\Psi \, \pa_a ~+~ i\, \fracm 1{\,16} \, (\g_a)^{\a \, \b} \, \Psi^{1/2} 
\, \Big( \, D_{\a} \Psi^{1/2} \, \Big) D_{\b}   ~~~.
}  \label{eq:04} \ee
If we vary the $\Psi$-superfield in (\ref{eq:04}) we find,
\be 
\d_S {\rm E}_{\a} ~=~ \fracm 12 L \, {\rm E}_{\a}  ~~~,~~~ \d_S 
{\rm E}_{a} ~=~ L \, {\rm E}_{a} ~+~  i\, \fracm 1{\,32} \, (\g_a)^{
\a \, \b} \, ({\rm E}_{\a} L) \, {\rm E}_{\b}  ~~~.
\label{eq:044} \ee
where $L \equiv (  {\Psi}^{-1} \d \Psi)$.  Interestingly enough, this 
same result is also obtained by varying $\Psi$ in (\ref{eq:03E}) and
(\ref{eq:03F}).  We are thus led to additional conclusions.  All the
remaining superfields ($\Psi_{[1]},~\cdots,~\Psi_{\[5\]}$ and ${\rm H}{
}_{\b}{}^m$) are superscale invariant and (\ref{eq:044}) represents 
a {\it {minimal}} 11D superspace scale transformation law.  

Let us use the second of the results in (\ref{eq:04}) to take a ``peek''
at some component fields in this conformal gauge.  Taking the limit as $\q \to
0$, we find
\be
{\rm e}_a {}^m (x) ~=~ ( \Psi {\big |}) \,\d_a {}^m ~~~,~~~ \psi_a
{}^{\b} (x) ~=~ i\, \fracm 1{\,32} \, (\g_a)^{\a \, \b} \, (D_{\a} \Psi
{\big |}) ~~~.
 \label{eq:041} \ee
Taking the determinant of both side of the first of these and the ``gamma'' 
trace of the second implies yields
\be
 \Psi {\big |} ~=~ \big[ \det ({\rm e}_a {}^m )\,\big]{\,}^{1/11}
     ~~~,~~~ D_{\a} \Psi {\big |} ~=~  i \fracm {32}{11} \,  
    (\g^a)_{\a \, \b} \,\psi_a {}^{\b}(x) ~~~. 
\label{eq:043} \ee
These equations show that within this gauge the determinant of the
component 11D vielbein and the ``gamma-trace'' of the component 11D
gravitino can reside in the single superfield $\Psi$.  The superfield
$\Psi$ has been often called  the conformal compensator. {\it {We
emphasize that all known off-shell constructions of supergravity possess
this superfield.}} Of course, since we have imposed no spinorial
differential constraints upon $\Psi$, there are $2^{31} - 1$ addition
bosonic d.f.\ and $2^{31}  - 32$ additional fermionic  d.f.\ that 
accompanying these fields. 

\section {11D Linearized Vielbein Semi-Prepotential  
Analysis: Anholonomy Coefficients}

~~~~In the limit of infinitesimal superfields in (\ref{eq:01}), we
``shift''
$\Psi$  according to $\Psi \to 1 + \Psi$ and expand the exponential in 
(\ref{eq:01}) to first order to find
\be
{\rm E}{}_{\a} ~=~ D_{\a} ~+~ \fracm 12 \, \Big( \D_{\a}{}^{\g} ~+~
    \Psi \d_{\a}{}^{\g} \Big) \, D_{\g}  ~+~ {\rm H}{}_{\a}{}^c 
    \, \pa_c  ~~~.
\label{eq:05} \ee
This form is convenient for considerations of infinitesimal supergravity 
fluctuations about a flat 11D background.   In this case, the
semi-prepotential\footnote{These are semi-prepotentials because they
must be subject to a complete set of (presently \newline ${~~\,~~}$
unknown) differential constraints.  Such constraints are discussed later
in this paper.  The \newline ${~~~~\,}$ true prepotentials are the
solutions  to the totality of such differential equations. } superfields 
$\Psi^{\[p\]}$  and ${\rm H}{}_{\a}{}^m$ are the appropriate variables
to consider for a  linearized analysis of 11D superfield supergravity. 
In this linearized  theory, the vectorial vielbein defined by
(\ref{eq:02}) and (\ref{eq:05})  takes the form, 
\be \eqalign{
{\rm E}{}_{a} \, =\,&  ~\pa_a \,+\,  i \, \fracm 1{\,32\,} \, \Big[ \, 
D_{\b} (\, \g_a \, \D \,)^{\b \, \g} ~+~  (D_{\b} \Psi )(\g_a )^{\b \g} 
\, \Big] \, D_{\g} \cr  
&+~ \Big[ ~ i\, \fracm 1{\,16\,} \, (\g_a)^{\a 
\, \b} \, ( \, D_{\a} {\rm H}{}_{\b}{}^c \, ) ~+~ \d_a {}^c \, \Psi ~-~ 
\Psi_a {}^c ~\Big] \, \pa_c ~~~.~~~
} \label{eq:06} \ee
By looking at the final line of this equation, we see that the 0-form and
2-form holonomy superfields are indeed the compensators for scale and 
Lorentz symmetries, respectively.    

We can use the linearized superframes to find the relations between these
linearized semi-prepotentials and the ``geometrical constraints'' to be 
imposed on the 11D superspace anholonomy.  To do this, we next compute
the graded commutator algebra of the linearized frames given in
(\ref{eq:05}) and (\ref{eq:06}).  Starting with the algebra of the
spinorial frames we find
\be \eqalign{ {~~~}
C_{\a \,\b }{}^{\e} &=~ \fracm 1{64} \, \Big[ ~ (\g^{\[2\]})_{\a \b} 
\, (\g_{\[2\]})^{\g \,\d}  \,-\, \fracm 1{60} \, (\g^{\[5\]})_{\a \b} 
\, (\g_{\[5\]})^{\g \,\d} ~ \Big] \, [ \, D_{\g} \D_{\d}{}^{\e} 
\,+\, \d_{\d}{}^{\e} (D_{\g} \Psi) \, ] ~~~,  
}\label{eq:13} \ee
\begin{equation}\eqalign{ {~~~}
C_{\a \,\b }{}^{c} &=~ ~i \,(\g^c)_{\a \b} ~-~  \, \fracm 1{32}
\,  (\g^{d e})_{\a \b} \, [ \,   (\g_{d e})^{\g \,\d} \,
D_{\g}{\rm H}{}_{\d}{}^{c} \, +\, 32  \d_d{}^c \Psi_e \,
- \, 16  \Psi^c{}_{d e}  \,]   \cr
&~~~~~+~ i \fracm 1{1,920}
\,  (\g^{d_1\cdots d_5})_{\a \b} \, [ \,  
     i (\g_{d_1\cdots d_5})^{\g \,\d} \,
D_{\g}{\rm H}{}_{\d}{}^{c} \,
+ \, 40 \d_{d_1}{}^c \Psi_{d_2\cdots d_5} \,  \cr
&~~~~~ ~~~~~ ~~~~~ ~~~~~ ~~~~~ ~~~~~ ~~~~~ -  \, \fracm 2{15}
\e{\,}^c{}_{ d_1\cdots d_5}{}^{\[5\]} \Psi_{\[5\]} \,]    ~~~,
}\label{eq:13A} \end{equation}
where we have utilized one of the Fierz identities noted in \cite{G6} and
as well the definition of $\D_{\a}{}^{\b}$ given in (\ref{eq:01}) in the 
latter of these two equations.  In reaching these results, we see that
neither spinorial nor bosonic differential restrictions have been
imposed upon any of the superfields that appear in (\ref{eq:05}).

We continue by calculating the commutator algebra of ${\rm E}_{\a}$ and
${\rm E}_{b}$ which leads to the anholonomy coefficients,
\be \eqalign{ {~~}
C_{\a \,b }{}^{\g} &=~ i \, \fracm 1{32} \, [ ~ D_{\a} D_{\d} 
     \,(\g_b \D)^{
\, \d \g}  \, ] ~-~ \fracm 12 \pa_b \D_{\a}{}^{\g} 
     ~-~ \fracm 12 ( \pa_b \Psi ) \,
\d_{\a}{}^{\g}  ~+~ i \, \fracm 1{32} 
     \, [ ~ D_{\a} D_{\d} \Psi \, ] \, (\g_b)^{\,
\d \g} ~~~, \cr  
C_{\a \, b }{}^{c} &=~ i \, \fracm 1{16} \, 
     (\g_b)^{\,\b \g}\, [ ~ D_{\a} 
D_{\b} {\rm H}{}_{\g}{}^{c}  \, ] 
    ~-~ (\, \pa_b {\rm H}{}_{\a}{}^{c} \,
)   ~+~  \fracm 1{32} \, [ ~ D_{\b} \,(\g_b \D \g^c \,)^{\,\b}{}_{ \a} 
\, ] ~-~ D_{\a} \Psi_b {}^c  \cr 
&~~~~+\, [\, (D_{\a}\Psi) \, \d_b {}^c ~-~ \fracm 1{32} 
( \g^c \g_b)_{\a} {}^{\g} \, (D_{\g}\Psi)  \, ] ~~~.
}\label{eq:14} \ee
Finally, we calculate the commutator algebra of two vectorial frames
which leads to the anholonomy coefficients,
\be \eqalign{ {~~~~~~~}
C_{a \, b }{}^{\g} &=~  i\, \fracm 1{\,32\,}
     \, [ ~ - D_{\b} (\g_{\[ a|}
\pa_{|b \]} \, \D)^{\, \b \g} ~+~ (D_{\b} \pa_{\[a|} \Psi)\, 
    (\g_{|b \]})^{\, \b \g}   ~ ] ~~~, \cr 
C_{a \,b }{}^{c} &=~ - \, i\, \fracm 1{\,16\,} 
     \, (\g_{\[a|})^{\a \, \b} \,[ \,
\pa_{|b \]} D_{\a} {\rm H}{}_{\b}{}^c \, ] 
    ~-~ (\pa_{\[ a |} \,\Psi_{|b \]}  {}^c ) 
~-~ \d_{\[a|} {}^c \, (\pa_{| b \] } \Psi)  ~~~.
}\label{eq:15} \ee

After the spin-connections are added to the dimension 1/2 anholonomy
coefficients, we can choose to impose the conditions (see also
\cite{G80} eq.(16))
\be \eqalign{ {~~} T_{\a \,\b }{}^{\g} &=~ \dots ~+~ \fracm 12 \d_{(
\a}{}^{\g}  {\cal J}_{\b)} ~+~
\fracm 1{32} \,  (\g^{a})_{\a \b} \, (\g_{a})^{\g \d} \, {\cal J}_{\d}
~~~, \cr T_{\a \,b }{}^{c} &=~ \dots ~+~ {\cal J}_{\a}\, \d_b {}^c ~-~
\fracm 1{32}  (\g^c \g_b)_{\a} {}^{\g} \, {\cal J}_{\g}    ~~~.
}\label{eq:T1} \ee {\it{This is the set of constraints that should
replace our d = 1/2 torsion set given in 1996 as a partial off-shell
construction.  The first explicit term  in each equation above was
absent in our previous work on partial off-shell superspace.}}

\section{How About Howe's 1997 Theorem}

~~~~In his 1997 work, Howe introduced ``Weyl superspace,'' which
necessarily  treats the scale transformation covariantly, so that the
scale transformations  of his vielbein (which we denote by ${\Tilde {\rm
E}}{}_A$ in the following to distinguish it from the usual frames) take 
the forms,
\be 
\d_S {\Tilde {\rm E}}{}_{\a} ~=~ \fracm 12 L \, {\Tilde {\rm E}}{}_{\a}  
~~~,~~~ \d_S {\Tilde {\rm E}}{}_{a} ~=~ L \, {\Tilde {\rm E}}{}_{a} ~~~.
\label{eq:041} \ee
Thus the first point to note about Howe's Weyl superspace, is that the
vectorial frame superfields (after imposing the condition $T_{\a \b}{}^c
= i (\g^c)_{\a \b}$) must contain more degrees of freedom than the
minimal vectorial frame in (\ref{eq:02}).  This is confirmed by the
observation\footnote{We  thank P.\ Howe for bringing to our attention
the existence of this frame.} that another vectorial frame (${\Bar {\rm
E}}{}_{a}$) defined by
\be
{\Bar {\rm E}}{}_{a} ~\equiv~ {\Tilde {\rm E}}{}_{a} ~+~ i\,
\fracm 1{\,264\,} \, (\g_a)^{\a \, \b} \, {\Tilde C}{}_{\a \g}{
}^{\g}\, {\Tilde {\rm E}}{}_{\b} 
\label{eq:042} \ee
possesses the exact same transformation law as the vectorial frame in
(\ref{eq:04}).  The second term here removes the extra degrees of freedom
contained in the first.  This implies that Howe's vectorial frame has
all the degrees of freedom of his spinorial frame {\it {plus}} additional 
degrees of freedom contained in a unconstrained spinor superfield.  
Therefore, his Weyl superspace introduces an additional 64 d${}_S$ 
degrees of freedom compared to the minimal set of frame superfields
defined by (\ref{eq:01}) and (\ref{eq:02}).

In another part of his paper, a process we have long called
``degauging''  is carried out.  Here, he finds a dimension 1/2 spin-1/2
quantity denoted by
$K_{\a}$.  However, he ultimately dispenses with this quantity with a
remark, ``...and recognize that it can be removed by a super-Weyl
transformation.''   This encapsulates our fundamental disagreement with
Howe.

We expect that when the totality of M-theory corrections are found, some 
of them will necessarily activate a spinorial auxiliary field multiplet
of dimension 1/2. Our reason for this is somewhat subtle.  In some of
our first  reports on how string corrections modify low-energy effective
actions and  consequently superspace geometry \cite{G1,G2} we pointed
out that  there are in fact {\it {two}} {\it {distinct}} sources of
corrections.   One occurs by integrating out the higher massive mode
corrections to  the zero mass field equations and the other occurs from
the actual  quantum loop corrections to the theory.  We noted that there
is a way  to distinguish between these two types of corrections. 
Namely, one type of correction breaks scale invariance and the other
does not.  We also observed that under duality transformations, the
roles of the two types  of corrections are exchanged.

Now let us ask what happens in the context of known off-shell
formulations of superfield supergravity when the supergravity system 
is coupled to a non-scale invariant system?  The answer is well known, 
the auxiliary fields within the superspace scale compensator become 
``active.''   All of our investigation of the 11D fluctuations suggest 
that 11D supergravity has the usual canonical structure observed 
previously in other prepotential formulations of supergravity theories.  

>From component level results, the first non-trivial bosonic correction 
to 11D supergravity/M-theory is expected to be of the form\footnote{We 
warn the reader that we only mean this equation {\it {symbolically}}.  
In fact, the full uncon- \newline ${~~~~~}$ tracted Riemann curvature 
tensor appears. The quantity $X_{CS}^{ LL~(7)}$ is the Lorentz Chern-
\newline ${~~~~~}$ Simons 7-form.} 
\be
{\cal L}_{(1)} ~\sim~ (\ell_{11})^6 ~[ \, {\rm e}^{-1} \, R^4 ~+~
F^{(4)} \, \wedge \, X_{CS}^{ LL~(7)} ~] 
\label{eq:042} ~~~, \ee
which is {\it {not}} scale-invariant since it possesses a weight of -3. 
Above we have introduced a parameter denoted by $\ell_{11}$ of dimensions
(mass)${}^{-1}$.  We are therefore led to the expectation that since 
$ {\cal L}_{(1)}$ is scale non-invariant, it will activate the
auxiliary fields in the 11D conformal compensator, $\Psi$.  

By the same token, scale-invariant  (in the sense defined in \cite{G1,G2})
corrections will not activate this multiplet and it {\it {only}} then 
becomes irrelevant.  Since the correction above is expected to simply be
the lowest order one, there should occur higher order terms also.  It is
perhaps useful to discuss an example of such a higher order correction
that is scale-invariant. One such weight zero Lagrangian is given by
\be
{\cal L}^* ~\sim~ (\ell_{11})^9 \, {\rm e}^{-1} \,F^{(4)} \, R^5
\label{eq:042ZZ} ~~~. \ee
Should such an operator occur in the M-theory effective action, it is our
conjecture that it {\it {cannot}} activate the spin-1/2 multiplet of currents.
This is three orders higher than the lowest order correction.
So it will be difficult to check this conjecture.

\section {4D, $N$ = 8 $\leftrightarrow$ 11D, N = 1 Oxidation/Reduction?}

~~~~Howe's critique of our work also ignored something else that we
stated in our discussion of the partial off-shell description of 11D
supergravity.  Namely we were very much aware that other tensors (besides
${\cal J}_{\a}$) seem required to describe a completely off-shell theory.
Our concluding remarks explicitly say that we were aware of at least
two arguments for the existence of these other tensors.  We think it is
now useful to especially and explicitly review one of these arguments.

Many years ago \cite{GG} a much overlooked result was derived within the
context of 4D, $N$ = 8 supergravity in superspace.  A study was made of
what (if any) unusual space-time differential constraints would be
imposed upon the Weyl tensor by assuming the traditional sets of
constraints 
\be 
T_{ \a i \, \b j}{}^{c} ~=~ 0 ~~~,~~~ T_{ \a i \, \Dot \b}{}^{~ j ~ c} 
~=~ i \, \d_{\a}{}^{\g} \, \d_{\Dot \b }{}^{ \Dot \g} \,  \d_i {}^j ~~~,
\label{eq:043X} \ee
used on lower $N \le 4 $ superspace supergravity theories.  In
distinction to all lower dimensional theories it was found that these
conditions alone were sufficient to impose within the linearized
approximation
\be
\pa^{\un a } \, \pa^{\un b} \, w_{\a \, \b \, \g \d} ~=~ 0 ~~~,
\label{eq:043Y} \ee
as a restriction on the linearized Weyl tensor. This is symptomatic of
acausal propagation, so we concluded that to avoid this above the $N$ = 4
barrier, it would be necessary to replace (\ref{eq:043X}) by equations
\be 
T_{ \a i \, \b j}{}^{c} ~\neq~ 0 ~~~,~~~ T_{ \a i \, \Dot \b}{}^{~ j ~ c} 
~\ne~ i \, \d_{\a}{}^{\g} \, \d_{\Dot \b }{}^{ \Dot \g} \,  \d_i {}^j ~~~.
\label{eq:043Z} \ee
If we ``oxidize'' this result to eleven dimensions, it seems to nicely
match the results of a more recent work \cite{CGNN}.

The primary assertion in the work of \cite{HOW} is the notion that {\it
{all}} the equations of motion for 11D supergravity/M-theory follow from
the condition $T_{\a \b}{}^c = i (\g^c)_{\a \b}$.  This has been
referred  to as ``Howe's Theorem'' (e.g., \cite{CGNN,PVW}).  Let us,
however, return  to the result in (\ref{eq:13A}).   Using the notation of
\cite{CGNN} we can write (in our conventions)
\be
T_{\a \b}{}^c ~=~ i \,(\g^c)_{\a \b} ~+~  \,\fracm 12 \,  (\g^{
\[2\]})_{\a \b} \, X_{\[2\]}{}^c ~ +~ i\, \fracm 1{120} \, 
(\g^{\[5\]})_{\a \b} \, X_{\[5\]}{}^c  
\label{eq:043} ~~~. \ee
If we perform a dimensional compactification on a torus, this leads to
(\ref{eq:043Z}).  Thus, it was our expectation that some new field
strengths could appear in the dimension zero superspace
torsions.

Upon comparing this with (\ref{eq:13A}), it is seen that both real field 
strengths $X_{\[2\]}{}^c$ and $X_{\[5\]}{}^c$ are dependent {\it
{algebraically}} on some of the holonomy $p$-forms, but are
independent of $\Psi$ and $\Psi_{\[2\]}$:
\be\eqalign{ ~~~ 
X_{a b}{}^c & ~=~  \fracm1{16} \Big[\,  (\g_{a b})^{\g \,\d} \, 
D_{\g}{\rm H}{}_{\d}{}^{c} \, +\, 16 \d_{\[a}{}^c \Psi_{b \]} \, 
- \, 16 \Psi_{a b}{}^c \, \Big]  ~~~, \cr  
~~~ X_{a_1\cdots a_5}{}^c & ~=~ \fracm1{16} \Big[\, 
     i (\g_{a_1\cdots a_5})^{\g \,\d} \, D_{\g}{\rm H}{}_{\d}{}^{c} \, +
\, \fracm{1}{3} \d_{\[ a_1}{}^c \Psi_{a_2\cdots a_5 \]} \, 
      -  \, \fracm 2{15}
\e{\,}^c{}_{ a_1\cdots a_5}{}^{\[5\]} \Psi_{\[5\]} \, \Big] ~~~.     
\label{X's}} \ee 
The vanishing of parts of the
$X$-field strengths simply  determine $\Psi_{\[1\]}$,
$\Psi_{\[3\]}$, $\Psi_{\[4\]}$ and
$\Psi_{\[5\]}$ in terms  of H${}_{\a}{}^b$.  Therefore the solution to
these algebraic equations allow us to  impose the further conventional
constraints on the $X$-tensors
\be 
X_{a b}{}^b ~=~ 0 ~~~, ~~~ X_{\[ a b c \]} ~=~ 0 ~~~,~~~ X_{a_1 \dots
a_4 a_5}{}^{a_5} ~=~ 0 ~~~, ~~~ X_{\[ a_1 \dots a_5 b \]} ~=~ 0 ~~~.
\label{eq:043A} 
\ee 
This leaves only $\Psi$ and H${}_{\a}{}^b$
with\footnote{The quantity $\Psi_{
\[2\]}$ is the Lorentz compensator and is thus a pure gauge degree of
freedom.}  706 d${}_S$ degrees of freedom as the true variables that
describe the 11D supergravity theory.  Here H${}_{\a} {}^m$ is the gauge 
field for the Weyl theory degrees of freedom and $\Psi$ is the Goldstone
superfield for breaking superconformal symmetry to super Poincar\' e
symmetry.  In obtaining this result, we note that {\it {neither}} spinorial
nor bosonic differential constraints are imposed upon $\Psi$, $\Psi_{\[2\]}$ 
nor H${}_{\a}{}^b$ due the constraints listed in table three.  Finally the
authors of \cite{CGNN} impose the condition that $X_{\[2\]}{}^c = 0$ and this
is seen to be a genuine restriction on H${}_{\a}{}^b$.  While it is not
clear to us why this is preferable to the opposite $X$-restriction, it
is clear that the opposite choice $X_{\[5\]} {}^c = 0$ leads to a
smaller supergravity multiplet if it is viable.

Above we noted that the $X$-tensor superfield strengths are independent of 
$\Psi$.  In particular, this means that imposing conditions on them {\it
{cannot}} impose any conditions on $\Psi$.  This is a direct contradiction
of the main result of Howe's 1997 Theorem because we see that even if 
~$T_{\a\b}{}^c = i (\g^c)_{\a \b}$, no equations at all are imposed on
$\Psi$.  Looking back at the full non-linear expressions for $X_{\[2\]}
{}^c$ and $X_{\[5\]} {}^c$ (given in (\ref{eq:03HA}) and (\ref{eq:03HC})), 
these are also seen to be independent of $\Psi$, so the argument valid
for the linearized theory generalizes to the non-linear theory and
disproves Howe's assertion in general.

\section { A Simple Component-Field Modification}

~~~~At the level of component fields, the manner in which the spin-1/2
multiplet will manifest itself follows from the general discussion given
in ``{\it {Superspace}}'' \cite{ConSTRT} (p.\ 323) which is easily
generalized to 11D.  The transformation law of the graviton using only
the constraint given in (\ref{eq:02}) can be {\it {derived}} to be of
the form
\be  \eqalign{ {~~~~~~}
&\d_Q {\rm e}_a {}^m ~=~ -\, \e^{\b} \, [~ T_{\b a}{}^{b} ~+~ \psi_a 
{}^{\g} \, T_{\b \, \g}{}^b ~] \,  {\rm e}_b {}^m  ~~~, \cr
&\to~ \d_Q {\rm e}{}^{-1} ~=~ {\rm e}{}^{-1} \, \e^{\b} \, [~ i \, 
(\g^a)_{\b \g} \, \psi_a {}^{\g} ~+~ \fracm {341}{32}{\cal J}_{\b} 
~+~ \fracm 12 \, X_{\[2\]}{}^a \, (\g^{\[2\]})_{\b \g} \, \psi_a
{}^{\g} ~  \cr 
&{~~~~~~~~~~~~~~~~~~~~~~~~~~~~}+~ i\, \fracm 1{120} \, X_{\[5\]}{}^a
\, (\g^{\[5\]})_{\b \g} \, \psi_a {}^{\g}  ~] ~~~,  }
\label{eq:042H} \ee   
and in the presence of a spin-1/2 dimension 1/2 auxiliary field, the
${\cal J }_{\b}$ term is non-zero.  This has a very dramatic effect. 
The ${\cal J}_{\b}$ term is fermionic {\it {but}} supercovariant.  In
the on-shell theory there is only one perturbatively supercovariant
spinorial quantity upon which it can depend, the field strength of the
gravitino.  

The condition for a fermionic spin-1/2 dimension 1/2 auxiliary field 
to occur is that in the presence of the M-theory  corrections, the 
elfbein density  contains a purely supercovariant term in its supersymmetry
transformation  law.  In turn this means testing for the presence or 
absence of the spin-1/2 dimension 1/2 auxiliary field is simpler than 
one might imagine.  We only need a consistent procedure to find the 
elfbein 
transformation law with its M-theory corrections.  Finally, this 
equation (\ref{eq:042H}) shows how the ``$X$-tensors'' also modify the
on-shell graviton transformation laws.  The work in \cite{PVW} should
definitively lead to an answer as to what is the form of the transformation
law chosen by the lowest order correction in supergravity/M-theory.  

Let us now discuss what Peeters, Vanhove and Westerberg \cite{PVW} have 
presented.  Before doing this, it is perhaps useful to warn the reader
that due to their non-canonical field definitions, there are numbers of
re-definitions necessary in order to compare their results with canonical
ones.   As an example of what we mean by ``non-canonical field definitions,''
their equation (3.36) is a useful reference point.  It is clear that
by making a redefinition of their gravitino field, all of the terms
involving the space-time derivative of the local supersymmetry parameter 
can be removed from the rhs of this equation.  The use of an ordinary 
superspace formalism is simplest after implementing such re-definitions.  
We have not implemented such re-definitions, so instead 
of directly commenting on their results we will use them to {\it {motivate}} 
the appearance of certain superspace structures useful for future
study. 

Working in the basis of their equation (3.43) and multiplying it by 
an elfbein in order to form a trace, we find it suggestive of a
${\cal J}$-tensor of the form
\be
{\cal J}_{\b} \, \approx \,  i a_0 \,  (\ell_{
11})^6 \, t_8^{\[8^{\prime}\]} \, \eta_{a_5^{\prime} \[ a_5 |} \eta_{
| a \] a_6^{\prime}} \, (\g^{\[5\]})_{\b \g} \, W_{ a_1 a_2 \, a_1^{
\prime}  a_2^{\prime}} \, W_{a_3 a_4  \, a_3^{\prime} a_4^{\prime}}
\, \Big( \nabla^a T_{a_7^{\prime} a_8^{\prime}}{}^{\g} \, \Big) ~~,
\label{eq:042L} \ee
where $t_8^{\[8^{\prime}\]}$ is a very well known rank-8 tensor (see 
\cite{PVW} for a definition) and $a_0$ is some normalization constant.
{\it {As per our expectations, the non-scale invariant M-theory
correction would then have excited the spin-1/2 multiplet of currents.}}

>From our perspective, there is one other very puzzling result of their 
analysis.  Near the end of the last chapter, we noted that Cederwall, Gran,
Nielsen and Nilsson \cite{CGNN} chose $X_{\[2\]}{}^a = 0$.  We noted that
the opposite choice $X_{\[5\]}{}^a = 0$ would naturally lead to a smaller
supergravity multiplet.  Interestingly, Peeters, Vanhove and Westerberg
report from their calculation of the commutator algebra (within their
approximations) that they find $X_{\[5\]}{}^a = 0$.  As before, if we 
look at equation (3.39) of their work and compare it to our equation
(\ref{eq:042H}) it is suggestive of
\be
X_{\[5\]} {}^a
~\approx~ - \, a_1 \, (\ell_{11})^6 \, t_8^{\[8^{
\prime}\]} \,  W_{\[ a_1 a_2 |\, a_1^{\prime} a_2^{\prime}} \,   W_{ | a_3
a_4 |\, a_3^{\prime}  a_4^{\prime}} \, W_{|a_5 \] a_6 \, a_5^{\prime}
a_6^{\prime}}  \, \d_{a_7^{\prime}} {}^{\[ a_6} \d_{a_8^{\prime}} {}^{a
\]} ~~~.
\label{eq:042M} \ee
where $a_1$ is a second normalization constant. In their final basis they
find $a_0 = a_1 =0$  \cite{PVW}.  It is a problem for the future to study 
these terms in a complete supergeometry for various values of the parameters
and as well as adding additional terms involving the 4-form field strength 
to both the ${\cal J}$ and $X$ currents.  Let us also close by emphasizing 
that the results in  (\ref{eq:042L}) and (\ref{eq:042M}) are simply motivated
from \cite{PVW} and we have not independently checked their consistency 
{\it {via}} use of the superspace Bianchi identities.  They do mark, however,
the  {\it {first}} {\it {explicit}} equations\footnote{For example, the work 
of \cite{CGNN} does {\it {not}} relate the $X_{\[5\]}{}^a$ tensor to the  
field strength superfields \newline ${~~~~\,}$ of the conventional on-shell 
11D theory.} for the lowest order supergravity/M-theory currents given in
superspace.  This is exactly analogous to the initial presentation of the
Yang-Mills/open-string superspace current given in equation (8.3) of the 
first work in \cite{G2}.

\section {Off-shell 11D Supergravity/M-theory Tensors}

~~~~We have repeatedly referred to the 4D, $N$ = 1 supergravity theory as
providing a paradigm for the actual superspace geometrical structure required
for the M-theory effective action.  Perhaps this point is best illustrated by
more detailed comments.  It is a demonstrable fact that the following set of
{\em {solely}} conventional constraints\footnote{The critical reader may note
that this set of constraints, though equivalent, is slightly different 
\newline ${~~~~~\,}$ from that which implictly appears in our previous
chapters.} 
\vspace{0.01cm}
\begin{center}
\footnotesize
\begin{tabular}{|c|c|c|}\hline
$~~{\rm {Superfield}} ~{\rm {Determined}}~~$ & 
${\rm {Geometrical}}~{\rm {Constraint}}$ 
\\ \hline
$ ~~ {\rm E}_a {}^m ~~$ &  $ ~  i \,(\g_a)^{\a \b} \, T_{\a \b}{
}^b ~=~ 32 \d_a \, {}^b ~$ \\ \hline
$ ~~ {\rm E}_a {}^{\m} ~~$ & $ ~ (\g_a)^{\a \b} \, T_{\a \b}{}^{
\g} ~=~ 0 ~\,~$ \\  \hline
$ ~~~ \o_{\a ~ d e}~~$ & $~\,~~~~~ T_{\a \, [ d e]} ~-~ \fracm 2{55} 
\, (\g_{d e})_{\a}{}^{\g}  \, T_{\g b}{}^b ~=~0~\,~~~~~~~$ 
\\ \hline 
$ ~~  \o_{a ~ d e}~~$ & $ ~(\g_a)^{\a \b} \, R_{\a \b}{}^{d e} 
~=~ 0~$  \\  \hline
$ ~~ \Psi^{\[1\]}~~$ & $ ~(\g_{a b})^{\a \b} \, T_{\a \b}{}^{b} 
~=~ 0 ~$  \\ \hline
$ ~~ \Psi^{\[3\]}~~$ & $ ~(\g_{[a b|})^{\a \b} \, T_{\a \b}{}_{
| c]} ~=~ 0 ~$  \\ \hline
$ ~~ \Psi^{\[4\]}~~$ & $ ~(\g_{a b c d e})^{\a \b} \, T_{\a \b}{
}^{e} ~=~ 0 ~$  \\ \hline
$ ~~ \Psi^{\[5\]}~~$ & $ ~\fracm 1{\,6! \,} \, \e_{\[5\]}{}^{a b 
c d e f} \, (\g_{a b c d e})^{\a \b} \, T_{\a \b
}{}_{f} ~=~ 0 ~$  \\ \hline
\end{tabular}
\end{center}
\begin{center}
{Table 3: A Set of 11D SG/M-theory Conventional Constraints}
\end{center}
determine all the 11D supergeometry in terms of the semi-prepotentials 
H${}_{\a} {}^m$ and $\Psi$.   These constraints are such that only 
four independent superfields ${\cal W}_{a b c d}$, $X_{\[ a b \]}{}^c$,
$X_{\[ a_1\cdots a_5 \]}{}^c$ and ${\cal J}_{\a}$ are required to describe 
all the gauge-independent parts of the semi-prepotentials. The definitions
of these fields strengths are
\be \eqalign{ {~~~}
{\cal W}_{a b c d} &\equiv~ \fracm 1{32}\, \Big[\, i \,
( \g^e \g_{a b c d})_{\g}{}^{\a} \, T_{\a \, e}{}^{\g}
~-~  \fracm 13 \, (\g_{a b c d})^{\a \b}  ( \,\nabla_{\a}
T_{\b e}{\,}^e \,+\, \fracm {14}{5 \cdot 363} \, T_{\a k}
{\,}^k \, \, T_{\b l}{\,}^l  \,) ~ \Big] ~~~, \cr
X_{\[ a b \]}{}^c &\equiv~ \fracm 1{32}\, (\g_{a b})^{\a
\b} \, T_{\a \b}{}^c  ~~~, ~~~ X_{\[ a_1\cdots a_5 \]}{}^c
~\equiv~ i \, \fracm 1{32}\, (\g_{a_1 \cdots a_5})^{\a \b}
\, T_{\a \b}{}^c   ~~~, \cr {\cal J}_{\a}
&\equiv~ \fracm {4}{33} \, T_{\a b}{}^b  ~~~.}
\label{eq:01Z} \ee
Given these field strengths, we can discuss the equations of motion
associated with the 11D semi-prepotentials.  In the ordinary on-shell 
theory, ${\cal J}_{\a} = X_{\[a b \]}{}^c = X_{\[  a b d e f \]}{}^c = 
0$.  When general scale non-invariant M-theory corrections are described 
we must have ${\cal J}_{\a} \ne 0~ {\rm {and}}~ X \ne 0$.  If there are
scale-invariant M-theory corrections, then ${\cal J}_{
\a} = 0~ {\rm {and}}~ X \ne 0$.  Since the $X$-superfields are 
$\Psi$-independent they don't impose any equations of motion upon $\Psi$.  Its
equation of motion arises solely from imposing conditions on ${\cal 
J}_{\a}$.

This behavior is exactly like that of 4D, $N$ = 1 non-minimal
supergravity, a system possessing three off-shell multiplets, ${\cal W}{
}_{\a \b \g}$, $G_{\un a}$ and $T_{\a}$.  The equations of the on-shell
theory are given by $T_{\a} = G_{\un a} = 0$. Coupling this supergravity
theory  to conformal matter only  excites one of its off-shell multiplets
($T_{\a} = 0$, $ G_{\un a} \ne 0$) and  coupling it to non-conformal 
matter excites both ($ T_{\a} \ne 0, ~  G_{\un a} \ne 0 $).  The known 
lowest order M-theory corrections are not scale-invariant.  Due to this 
and in analogy to the coupling of non-minimal 4D, $N$ = 1 supergravity to
non-conformal matter we expect that both the $X$-multiplet of currents as 
well as the ${\cal J}$-multiplet of currents to be excited at lowest order 
in the $\ell_{11}$-expansion of 11D superspace geometry.   It is inconsistent
to attempt to describe the non-minimal 4D, $N$ = 1 supergravity multiplet 
by setting $T_{\a} = 0$ as a constraint and this is the direct analog of
imposing Howe's Theorem.

We end the section with a warning.  It has long been known that supergravity
theories must also satisfy additional constraints \cite{GSW2} that go beyond
the merely conventional  ones such as those in table three.   At present the
explicit  form of these addition constraints  are {\em {not}} known 
(although the work of \cite{CGNN} suggests that $ X_{\[a b \] c} = 0$ for
example).  Clearly more detailed investigation is required.

\section{Summary Discussion }

~~~~It is therefore our position that while we inaccurately reported the 
embedding of our long awaited spin-1/2 current multiplet in our 1996
paper, the part of Howe's 1997 work on {\it {ordinary}} Poincar\' e 
superspace can also be seen to {\it {necessarily}} include such a multiplet 
exactly as we predicted in 1980, {\it {but}} he concludes that it is 
irrelevant.   We believe that neither of these papers \cite{G6,HOW} 
is completely correct nor completely incorrect as
explained  in the previous chapters.

In our equation (\ref{eq:042H}), we have given a simple component level 
criterion to test for the presence or absence of the dimension 1/2 multiplet 
of currents.  The spin-1/2 and dimension 1/2 multiplet of currents that 
we have long anticipated is very closely related the first spinorial 
derivative of the eleven dimensional scale compensator.  Thus scale
non-invariant corrections should activate this multiplet and the superspace
torsions and curvatures will respond by the appearance of a spinorial
multiplet of currents, and frustrate attempts to eliminate the spin-1/2
multiplet of currents from the geometry of 11D supergeometry.  We contend 
that the yet-to-be completed component level analysis of \cite{PVW} shows 
a clear potential to support this position.

Our analysis of the linearized supergravity fluctuations clarifies the 
circumstances of the applicability of the 1997 Howe Theorem (that posits 
all 11D supergravity equations follow {\em {solely}} from the condition 
$T_{\a \b}{}^c = i (\g^c)_{\a \b}$).  The only way Howe's Theorem can 
be true is if $\Psi$ is a pure gauge degree of freedom (like $\Psi_{\[2\]
}$) and in the absence of super-scale symmetry this is impossible.  One 
more interesting implication of our 11D result is that it also cast doubt 
on a much older version \cite{BPTFFP} of this very same assertion within 
the context of the heterotic string effective action.  A repeat of this 
type of analysis for 10D, $N$ = 1 superspace yields the same results.  
With the verification that the ${\cal J}_{\a}$-tensor is present to 
accommodate the string corrections in 11D, it would follow that such a 
tensor must be present for, at least, 10D, $N$ = IIA supergravity also 
where it must take the form ${\cal J}_{\a \, i}$.  Furthermore this 
increases the likelihood that such an object appears in the correction 
to type-I closed and heterotic string modified superspace geometries in 
the form of ${\cal J}_{\a}$ ({\it i.e.}, setting one of the isospin copies 
of the IIA theory to zero).  Thus in addition to the $A_{\[3\]}$-tensor 
found \cite{G1,G2} some time ago, there would be a ${\cal J}_{\a}$-tensor  
and possibly an $X$-tensor in these theories also.  In fact, the expressions 
in (\ref{eq:042L}) and (\ref{eq:042M}) can be easily be {\it {interpreted}}
within the confines of 10D, $N$ = 1 superspace and presumably describe 
the same correction to the heterotic and type-I closed effective actions.   
The appearance of new auxiliary fields  may well offer the way around a 
vexing conundrum \cite{RTZ}.

We end by noting our twenty-year old conjecture \cite{G80} now has
increased chances for validation at last.

\bigskip

\moveright0.5in\vbox{
{\it ``The geometry of space is associated 
with a mathematical group.''} \newline
\hbox{~~~~~} -- Felix Klein} 

${~~~}$\newline
\noindent
{\bf {Acknowledgment}} \\[.0005in] \newline \noindent
${~~~~}$We wish to acknowledge discussions with
K.\ Peeters, P.\ Vanhove and A.\ Westerberg 
that were instrumental in the return of our attention to this matter. 
We also wish to acknowledge interesting conversations with P.\
Howe and M.\ Cederwall.



\begin{thebibliography}{66}


\baselineskip 8.5pt 

\bibitem{G1}S.\ J.\ Gates, Jr.\ and H.\ Nishino, Phys.\ Lett.\ 
{\bf {173B}} (1986) 52.

\bibitem{G2}S.\ J.\ Gates, Jr. and S.\ I.\ Vashakidze, Nucl.\ 
Phys.\ {\bf {B291}} (1987) 173; S.\ J.\ Gates, Jr. and H.\ Nishino, 
Nucl.\ Phys.\ {\bf {B291}} (1987) 205. 

\bibitem{G3}S.\ Bellucci and S.\ J.\ Gates, Jr., Phys.\ Lett.\ 
{\bf {208B}} (1988) 456. 

\bibitem{G4} S.\ Bellucci, S.\ J.\ Gates, Jr.\ and D.\ Depireux, 
Phys.\ Lett.\ {\bf {238B}} (1990) 315; 
H.\ Nishino, Phys.\ Lett.\ {\bf 258} (1991) 104.

\bibitem{GNZ}M.\ T.\ Grisaru, H.\ Nishino and D.\ Zanon,
Phys.\ Lett.\ {\bf {206B}} (1988) 625; {\it idem}.\ Nucl.\ Phys.\ 
{\bf {B314}} (1989) 363.

\bibitem{G5}S.\ Bellucci, S.\ J.\ Gates, Jr.\ , B.\ Radak, P.\ 
Majumdar and S.\ Vashakidze, Mod. Phys. Lett.\ {\bf {A21}} (1989) 
1985.

\bibitem{BdR}E.\ A.\ Bergshoeff and M.\ de Roo, Nucl.\ Phys.\ {\bf 
{B328}} (1989) 439. 

\bibitem{G6}H.\ Nishino and S.\ J.\ Gates, Jr., Phys.\ Lett.\ 
{\bf {388B}} (1996) 504.

\bibitem{G80}S.\ J.\ Gates, Jr., Phys.\ Lett.\ {\bf {96B}} (1980) 
305.

\bibitem{HOW}P.\ Howe, Phys.\ Lett.\ {\bf {415B}} (1997) 149. 

\bibitem{GSW}P. Breitenlohner, Phys.\ Lett.\ {\bf {67B}} (1977) 49; {\it
idem}.\ Nucl.\ Phys.\ {\bf {B124}} (1977) 500; W.\ Siegel, Harvard
preprint {\bf {HUTP-77/A068}} (November, 1977) and Harvard preprint {\bf
{HUTP-77/A080}} (November, 1977); S.\ J.\ Gates, Jr.\ and J.\ A.\
Shapiro, Phys.\ Rev.\ {\bf  {D18}} (1978) 2768; W.\ Siegel and S.\ J.\
Gates, Jr., Nucl.\ Phys.\ {\bf  {B147}} (1979) 77; S.\ J.\ Gates, Jr.\
and M.\ Brown, Nucl.\ Phys.\ {\bf  {B165}} (1980)  445.

\bibitem{CGNN}M.\ Cederwall, U.\ Gran, M.\ Nielsen and B.\ Nilsson,
``Manifestly Supersymmetric M-Theory,'' G\" oteborg ITP preprint,
hep-th/007035.

\bibitem{GSWc}S.\ J.\ Gates, Jr., K.\ S.\ Stelle and P.\ C.\ West, Nucl.\
 Phys.\ {\bf {B169}} (1980) 347. 

\bibitem{ConSTRT}M.\ Brown and S.\ J.\ Gates, Jr., Annals of Physics, 
{\bf {122}}, No. 2 (1979) 443; S.\ J.\ Gates, Jr.\ and W.\ Siegel, 
Nucl.\ Phys.\ {\bf {B163}} (1980) 519; S.\ J.\ Gates, Jr., M.\ T.\ 
Grisaru, M.\ Ro\v cek and W.\ Siegel, ``Superspace or One Thousand 
and One Lessons in Supersymmetry," Benjamin Cummings (Addison-Wesley), 
Reading, MA (1983).

\bibitem{TASI} 
S.\ J.\ Gates, Jr., ``Basic Canon in D = 4, N=1 Superfield Theory: Five 
Primer Lectures,'' Lectures given at Theoretical Advanced Study Institute 
in Elementary Particle Physics (TASI 97) {\it {Supersymmetry, Supergravity 
and Supercolliders}}, ed.\ J.\ Bagger (World Scientific,  1999), Singapore,
hep-th/9809064. 


\bibitem{GG}S.\ J.\ Gates, Jr.\ and R.\ Grimm, Phys.\ Lett.\ {\bf {133B}} 
(1983) 192.

\bibitem{PVW}K.\ Peeters, P.\ Vanhove and A.\ Westerberg, 
``Supersymmetric $R^4$ Actions and \\ Quantum corrections to 
Superspace Torsion Constraints,'' hep-th/0010182; {\it
idem}.\ ``Supersymmetric Higher-derivative Actions in Ten and Eleven 
Dimensions, the Associated Superalgebras and Their Formulation  in
Superspace," hep-th/0010167. 

\bibitem{GSW2}S.\ J.\ Gates, Jr., K.\  S.\ Stelle and P.\ C.\
West, Nucl. Phys. {\bf {B169}} (1980) 347.

\bibitem{BPTFFP}L.\ Bonora, P.\ Pasti, and M.\ Tonin, Phys.\ Lett.\ 
{\bf {188B}} (1987) 335; S.\ Ferrara, P.\ Fr\' e and M.\ Porrati, Ann.\ 
Phys.\ {\bf {175}} (1987)  112.

\bibitem{RTZ}
L.\ Bonora, M.\ Bregola, R.\ D'Auria, P.\ Fre, K.\ Lechner, P.\ Pasti, I.\
Pesando, M.\ Raciti, F.\ Riva, M.\ Tonin and D.\ Zanon,
Phys.\ Lett.\ {\bf {B277}} (1992) 306.

\end{thebibliography}
\end{document}